\documentclass[aps,prl,a4paper, twocolumn, 10pt, showpacs]{revtex4-1} %
\usepackage{bbm, amsmath, amssymb, amsthm, bm, textcomp,graphicx,color}
\usepackage[utf8]{inputenc}
\usepackage[T1]{fontenc}
\usepackage[english]{babel}

\theoremstyle{definition}

\begin{document}

\title{Detecting large quantum Fisher information with finite measurement precision}

\author{Florian Fr\"owis$^{1}$, Pavel Sekatski$^{2}$, Wolfgang D\"ur$^{2}$}
\affiliation{$^1$ Group of Applied Physics, University of Geneva, 1211 Geneva, Switzerland\\
$^2$ Institut f\"ur Theoretische Physik, Universit\"at Innsbruck, Technikerstra\ss e 21a, 6020 Innsbruck, Austria}
\date{\today}

\begin{abstract}
We propose an experimentally accessible scheme to determine lower bounds on the quantum Fisher information (QFI), which ascertains multipartite entanglement or usefulness for quantum metrology. The scheme is based on comparing the measurement statistics of a state before and after a small unitary rotation. We argue that, in general, limited resolution of collective observables prevents the detection of large QFI. This can be overcome by performing an additional operation prior to the measurement. We illustrate the power of this protocol for present-day spin-squeezing experiments, where the same operation used for the preparation of the initial spin-squeezed state improves also the measurement precision and hence the lower bound on the QFI by two orders of magnitude. We also establish a connection to Leggett-Garg inequalities. We show how to simulate a variant of the inequalities with our protocol and demonstrate that a large QFI is necessary for their violation with coarse-grained detectors.
\end{abstract}

\maketitle

With the enhanced control in modern quantum experiments, it is now possible to coherently manipulate large numbers of microscopic objects like photons and atoms with unprecedented accuracy. However, these advances go hand in hand with new challenges. On the one hand, the discrimination of large quantum systems requires increased resolution, while detectors are generally imperfect and their resolution is limited. On the other hand, the number of parameters increases exponentially in number of particles or modes, which makes a complete tomographic characterization of the system unfeasible. It is therefore of uttermost importance to identify few key properties of the system and design simple measurement set-ups to determine them.

One such key property is the quantum Fisher information (QFI), which is essentially a measure of how fast a given state changes under a given evolution. The QFI is not only an indicator of how useful a quantum state is for quantum metrology \cite{Helstrom_Quantum_1976,*Holevo_Probabilistic_2011,Braunstein_Statistical_1994} (e.g., to determine an unknown parameter such as a frequency or a magnetic field \cite{Giovannetti_Quantum-Enhanced_2004}), but is also provides a lower bound on multipartite entanglement \cite{Pezze_Entanglement_2009,*Hyllus_Fisher_2012,*Toth_Multipartite_2012}. Furthermore, the QFI was proposed as a measure for macroscopicity of quantum systems \cite{Frowis_Measures_2012,Frowis_Linking_2015} and plays a role in other situations \cite{Zanardi_Quantum_2008,*Smerzi_Zeno_2012,Toth_Quantum_2014}.

Here we propose a simple and experimentally feasible scheme to determine lower bounds on the QFI. 
The scheme is based on the comparison of the measurement statistics of the generated state before and after a short but finite unitary evolution $U = \exp(-i H t)$ (e.g., a phase shift or an external electromagnetic field).
We argue why in general a large QFI can not be witnessed with limited detector resolution. We overcome this restriction by a supplementary operation prior to the measurement. As our main result, we find that this additional operation is of the same complexity as the preparation for relevant examples allowing to use the same experimental toolbox. As examples, we discuss squeezing in photonic and spin systems. For spin ensembles, we demonstrate that additional spin-squeezing operations \cite{Kitagawa_Squeezed_1993,Meyer_Experimental_2001,*Gross_Nonlinear_2010,*Hamley_Spin-nematic_2012,Riedel_Atom-chip-based_2010} allow one to improve bounds on the QFI by up to two orders of magnitude for realistic parameter values.

\begin{figure}[b!]
\centerline{\includegraphics[width=\columnwidth]{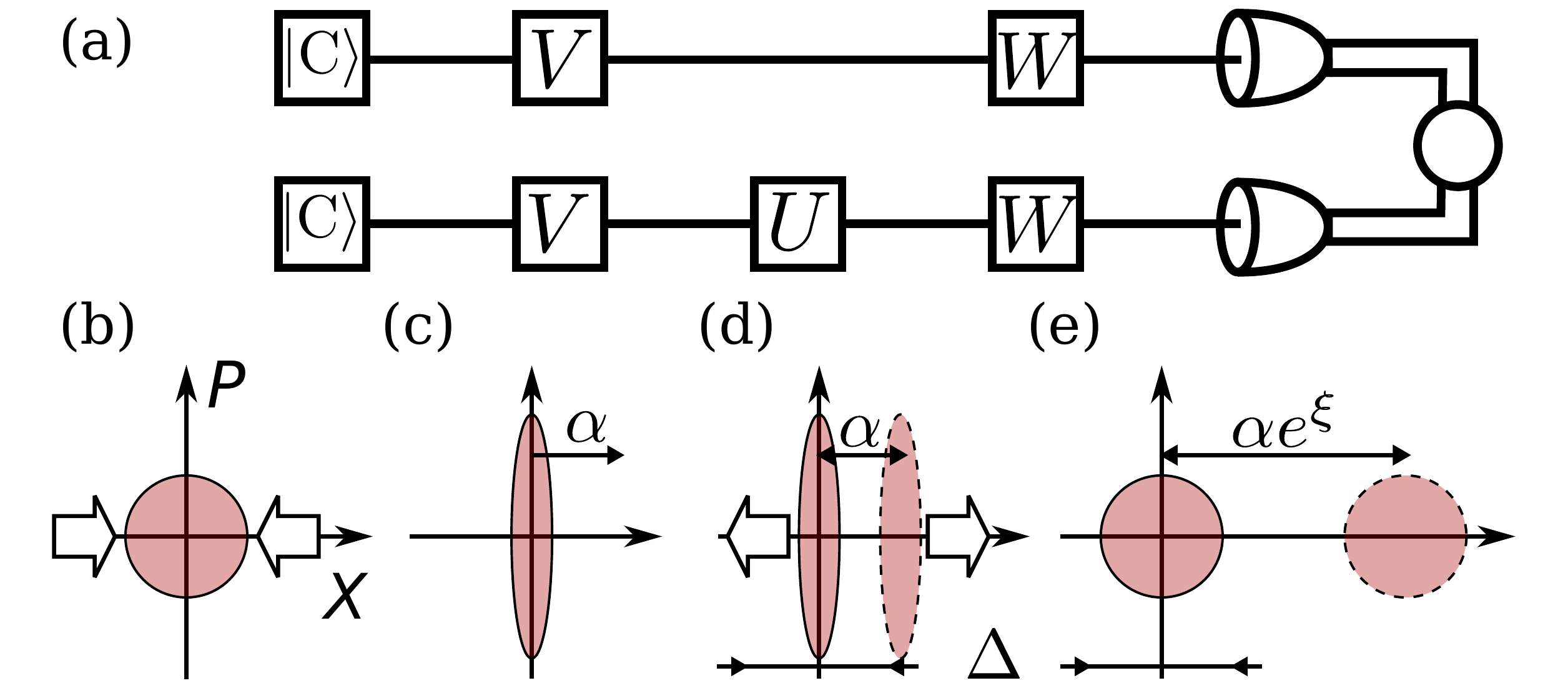}}
\caption[]{\label{fig:scheme}
  (a) Schematic of preparation, measurement and comparison of the two states used for the lower bound Eq.~(\ref{eq:3}). The unitary $V$ generates the nonclassical properties when applied to a classical state $| C \rangle $. $U$ is a linear rotation whose action on $\left| \phi_0 \right\rangle = V \left| \mathrm{C} \right\rangle $ potentially reveals the presence of large QFI. $W$ is of the same complexity as $V$ and helps to overcome limitations in the measurement resolution. (b-e) Photonic squeezing (see example 1 in the text). The shaded area corresponds to the phase space distribution of the state. 
  (b) Preparation of $| \mathrm{C} \rangle = \left| 0 \right\rangle $. The white arrows indicate the action of the squeezing operation $V$ with parameter $\xi$. (c) The action of $U$: The squeezed state is optionally displaced by $\alpha$, later illustrated with dashed contours. (d) Distinguishing the squeezed vacuum from the displaced squeezed vacuum requires a certain detection precision. Large coarse-graining $\Delta$ prevent this distinction. The white arrows indicate the action of the squeezing operation $W = V^{\dagger}$. (e) After the back-squeezing, it is possible to distinguish the states even with large $\Delta$.
}
\end{figure}

Furthermore, we establish a connection to Leggett-Garg inequalities \cite{Leggett_Quantum_1985} (more precisely the no-signaling in time (NSIT) variant \cite{Kofler_Condition_2013,Clemente_Necessary_2015}). As a consequence NSIT protocols can be simulated with our proposal. Interestingly, we find that a large QFI is necessary for their violation when dealing with coarse-grained detectors.

\textit{Lower bounds on QFI.---}
Even though the QFI is a nontrivial function of a state and $H$, tight lower bounds have been found \cite{Kholevo_Generalization_1974,*Hotta_Quantum_2004,*Pezze_Entanglement_2009,*Frowis_Tighter_2015,Frowis_Kind_2012,Strobel_Fisher_2014}. On the one hand, there exist bounds based on the Heisenberg uncertainty relation \cite{Kholevo_Generalization_1974,Hotta_Quantum_2004,Pezze_Entanglement_2009,Frowis_Tighter_2015}. On the other hand, one can bound the QFI by measuring the response of a state to an (infinitesimal short) application of a unitary evolution $U = \exp(-i H t)$ \cite{Frowis_Kind_2012,Strobel_Fisher_2014}. The latter also corresponds to the usage of the prepared state in a metrology scenario, where $H$ depends on some unknown parameter that shall be determined. It is well known that entangled states offer a (quadratic) improvement as compared to classical states \cite{Giovannetti_Quantum-Enhanced_2004}.

Here, we focus on the second approach and derive a lower bound on the QFI for finite $t$.
Given a unitarily evolving state $\rho(t)= \exp(-i H t) \rho(0) \exp(i H t)$, the QFI is a function of $\rho(0)$ and the generator $H$, but is independent of $t$ \cite{Braunstein_Statistical_1994}. With the fidelity $F(\rho,\sigma) = \mathrm{Tr}\sqrt{\sqrt{\rho}\sigma \sqrt{\rho}}$ of two states $\rho,\sigma$ and the Bures distance $d_{\mathrm{B}}(\rho,\sigma) = \sqrt{2} \left[ 1- F(\rho,\sigma) \right]^{1/2}$, the QFI $\mathcal{I}$ is implicitly defined via
\begin{equation}
\label{eq:1}
d_{\mathrm{B}}[\rho(t),\rho(t+dt)]^2 = \frac{1}{4} \mathcal{I}_{\rho}(H) dt^2.
\end{equation}
We refer to the literature for further properties and explicit formulas of $\mathcal{I}$ \cite{Braunstein_Statistical_1994,Toth_Quantum_2014}. From Eq.~(\ref{eq:1}), it is possible to derive the inequality
\begin{equation}
\label{eq:2}
F \left[ \rho(0),\rho(t) \right] \geq \cos \left[ \frac{1}{2} \sqrt{\mathcal{I}_{\rho}(H)}\,t \right],
\end{equation}
which is valid for $  |t| \leq \pi /\sqrt{\mathcal{I}_{\rho}(H)}$ \cite{Uhlmann_energy_1992,Frowis_Kind_2012}.
This implies that a large QFI is necessary for a rapid reduction of the quantum fidelity between $\rho(0)$ and $\rho(t)$. Consider now a generalized measurement $\left\{ \Omega_m \right\}_m$ with discrete outcomes $m$, $\Omega_m \geq 0$ and $\sum_m \Omega_m = \mathbbm{1}$. Let us denote the probabilities of measuring $m$ by $p_m = \mathrm{Tr} [\Omega_m \rho(0)]$ and $q_m = \mathrm{Tr} [\Omega_m \rho(t)]$, respectively. It holds that the Bhattacharyya coefficient $B_{\Omega} = \sum_m \sqrt{p_m q_m} \in [0,1]$ upper bounds the fidelity, that is, $B_{\Omega} \geq F[\rho(0),\rho(t)]$. There always exists an optimal measurement such that equality holds. Inserting this into Eq.~(\ref{eq:2}) and inverting it, we find the general inequality
\begin{equation}
\label{eq:3}
\mathcal{I}_{\rho}(H) \geq \frac{4}{t^2} \arccos^2 B_{\Omega},
\end{equation}
where for later we abbreviate $\mathcal{B} = 4 t^{-2} \arccos^2 B_{\Omega}$. 
While Eq.~(\ref{eq:3}) is valid for all $t$, it can only be saturated if $  |t| \leq \pi /\sqrt{\mathcal{I}_{\rho}(H)}$. The benefit of this inequality for experiments is obvious. For a given initial state $\rho(0)$, one chooses a small enough $t$ and generates $\rho(t)$. By measuring $\rho(0)$ and $\rho(t)$ with $\{\Omega_m\}_m$, one calculates $B_{\Omega}$ and directly gets a lower bound on $\mathcal{I}$.

The presented bound differs from other approaches \cite{Strobel_Fisher_2014}, where one has to scan $t$, fit the results and use the approximation of infinitely small time steps. Our bound does not rely on such fittings and approximations, which circumvents additional uncertainties coming from the fitting procedure. 
It also allows one to cope with statistical uncertainties. While $\mathcal{B}$ decreases with $t$ in general, this is not the case when a finite error bar $\delta$ on the estimated value of $B_\Omega$ is taken into account. Even in the perfect case, that is, $B_{\Omega} = \cos (\sqrt{\mathcal{I}}t/2)$, it is required that $t > 2 \arccos(1-\delta)/\sqrt{\mathcal{I}_{\rho}(H)}$ in order to find $B_{\Omega} + \delta < 1$ and hence to have a meaningful bound.

\textit{Typical experimental situations.---}
Equation (\ref{eq:3}) is a general bound on the QFI that is not restricted to any particular preparation, time evolution and measurement. We now focus on common experimental setups, in particular photons and spin ensembles. There, one typically has annihilation and creation operators. For photons, they are denoted by $a$ and $a^{\dagger}$, respectively, and follow the Heisenberg algebra $[a^{\dagger}, a] = \mathbbm{1}$. In spin systems, one has $S_{+}$ and $S_{-}$, which form together with $S_z = \frac{1}{2} [S_{+},S_{-}]$ a SU(2) algebra: $[S_z, S_{\pm}] = \pm S_{\pm}$. In the following, we call an operator linear or nonlinear with respect to polynomials of the creation and annihilation operators.

Let us consider a generic protocol for ensembles of $N$ spin-$1/2$ particles for detecting large QFI [see also Fig.~\ref{fig:scheme} (a)
(with $W = \mathbbm{1}$ for the moment)]. 
 The experiment starts by generating an initial state $| \mathrm{C} \rangle $, which is often a spin-coherent state. Here, we choose it to be oriented in $x$ direction, that is $S_x \left| \mathrm{C} \right\rangle = N/2 \left| \mathrm{C} \right\rangle $ with $S_x = \frac{1}{2}(S_{+} + S_{-})$. From this, one reaches the desired quantum state $| \phi_0 \rangle $ via a nontrivial operation $V$, that is, $| \phi_0 \rangle  = V \left| \mathrm{C} \right\rangle $. The simplest and most used operations are quadratic resulting in a two-body interaction. An example is the one-axis twisting
\cite{Kitagawa_Squeezed_1993}, which is nowadays routinely generated in many labs \cite{Meyer_Experimental_2001,*Gross_Nonlinear_2010,*Hamley_Spin-nematic_2012,Riedel_Atom-chip-based_2010}. It is defined as
\begin{equation}
\label{eq:4}
V_{\mu} = e^{-i \nu S_x} e^{-i \mu/2 S_z^2}.
\end{equation}
The first operation is the nontrivial step in the sense that it creates the nonclassical features of the state $| \mathrm{S}_{\mu} \rangle = V_{\mu} \left| \mathrm{C} \right\rangle $. For $\mu < O(N^{-1/2})$, spin-squeezing is generated, while $ \left| \mathrm{S}_{\mu} \right\rangle $ is called oversqueezed (or non-Gaussian) for larger $\mu$ values. The second operation in $V_{\mu}$ locally rotates the state and $\nu$ is fixed such that the variance of $S_z$ is maximal, while $S_y = \frac{1}{2i}(S_{+}-S_{-})$ has minimal variance (see Ref.~\cite{Kitagawa_Squeezed_1993} for the explicit formulas).

For $\mathcal{B}$, it is necessary to compare $| \phi_0 \rangle $ to $| \phi_1 \rangle = U \left|\phi_0  \right\rangle $. The Hamiltonian in $U$ is chosen to be $H = S_z$. It has to be linear if the QFI should provide any implications about entanglement. Note that $\left| \mathrm{C}\right\rangle $ has $\mathcal{I}_{\mathrm{C}}(S_z) = N$, which is maximal for all separable states. Hence, it is desirable to overcome this value. In the present system, the maximal value for all states is $\mathcal{I}(S_z) = N^2$ (reached by the GHZ state $\left| 0 \right\rangle ^{\otimes N} + \left| 1 \right\rangle ^{\otimes N}$). One-axis twisted states exhibit $\mathcal{I}_{|\mathrm{S}_{\mu}\rangle}(S_z) = O(N^{4/3})$ in the squeezed regime and up to $\mathcal{I}_{|\mathrm{S}_{\mu}\rangle}(S_z) \approx N^2/2$ in the oversqueezed regime.

The measurement is often restricted to be linear. In our case, these are collective observables that measure the total spin in a certain direction in space. For the given $H$, it is best to measure in the $x$-$y$ plane, that is, $\Omega_m = \left| m \right\rangle\!\left\langle m\right|  $ with $S_{\alpha}\left| m \right\rangle  = m \left| m \right\rangle $ for $S_{\alpha}=\frac{1}{2}(e^{i \alpha} S_{+} + e^{-i\alpha} S_{-}) $ and $m \in \{-N/2,\dots,N/2\}$.
As can be easily seen with numerous examples, quadratic operators for the state preparation $V$ and linear measurements $\Omega_m$ are fully sufficient to witness a large QFI with Eq.~(\ref{eq:3}).

\begin{figure}[t!]
\centerline{\includegraphics[width=\columnwidth]{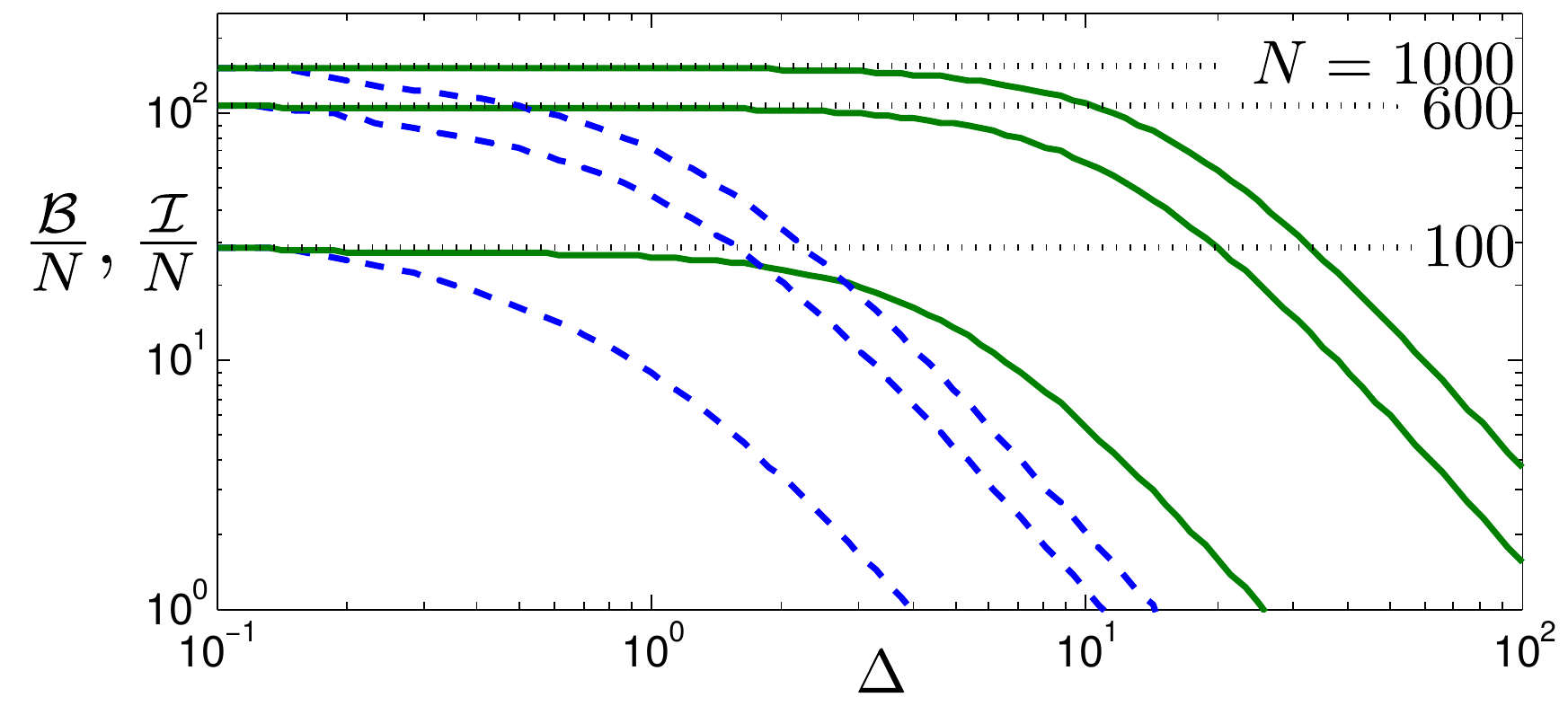}}
\caption[]{\label{fig:optsqueezing} Lower bounds $\mathcal{B}/N \leq \mathcal{I}/N$ for one-axis twisted states $| \mathrm{S}_{\mu} \rangle =V_{\mu} \left| \mathrm{C} \right\rangle $ obtained with coarse-grained collective observables for different $\Delta$ [see Eqs.~(\ref{eq:3}), (\ref{eq:4}) and (\ref{eq:5})]. For each $N = 100, 600, 1000$ (from bottom to top), the bounds are plotted for $W = \mathbbm{1}$ (blue dashed line) and $W = V^{\dagger}$ (green solid line) and compared to $\mathcal{I}/N$ (black dotted line). The values of $\mu$ in are chosen such that the variance in $y$ direction is minimal (see Ref.~\cite{Kitagawa_Squeezed_1993} for the explicit expressions). The improvement of the bounds through a nontrivial $W$ is particularly significant for $\Delta \approx 10$. The times chosen are $t = 10^{-2}$ for $N = 100$ and $t = 10^{-3}$ for $N = 600$ and $N = 1000$. Note that the results are optimized over the measurement axis in the $x$-$y$ plane.}
\end{figure}

\textit{Generic problem with measurement accuracy.---}
We now turn to more realistic descriptions of the measurement apparatus. As we will see, limited detector resolution in the spectrum of the observable generically leads to a drop of $\mathcal{B}$.
In some instances (such as the GHZ state), already the loss of the distinguishability between even and odd eigenstates results in trivial lower bounds on the QFI.
To study the effect of finite detection resolution, we use the following model of the coarse-grained detectors. Instead of $\Omega_m = \left| m \right\rangle\!\left\langle m\right| $, we consider a continuous measurement
\begin{equation}
\label{eq:5}
\Omega_\alpha(x) =
\frac{1}{\sqrt{2\pi} \Delta} e^{-(x-S_{\alpha})^2/(2 \Delta^2)}.
\end{equation}
Given a test state $| m \rangle $, $\left\langle m \right| \Omega(x)\left| m \right\rangle $ is normally distributed around $m$ with standard deviation $\Delta$.  The perfect measurement $\Omega(x) \rightarrow \delta(x-m) \left| m \right\rangle\!\left\langle m\right|$ is well approximated  for $\Delta\rightarrow 0$. In Fig.~\ref{fig:optsqueezing}, we plot $\mathcal{B}/N$ for the one-axis twisted states (blue dashed curve) for several hundreds of spins. We observe how sensitive the bound is to a rather small increase of $\Delta$. 

A heuristic argument suggests that a limited measurement resolution generically leads to a limited benefit of Eq.~(\ref{eq:3}) for $N\gg 1$.
We switch to the Heisenberg picture and consider the equivalent situation in which the state $| \phi_0 \rangle $ is observed through two different measurement bases $\left| m \right\rangle $ and $\left| \tilde{m} \right\rangle = U^{\dagger} \left| m \right\rangle $. The question of how well one is able to distinguish $| \phi_0 \rangle$ from $| \phi_1 \rangle $ is hence equivalent to how different $\left| m \right\rangle $ and $\left|\tilde{m}\right\rangle$ are when projected to the subspace spanned by $\left| \phi_0 \right\rangle$.
Suppose that one wants to show $\mathcal{I}_{\phi_0}(S_z) \geq O(N^x)$ with $1<x\leq 2$. 
For a tight bound, this necessarily implies $t = c N^{-x/2} \ll 1$ ($c$ a constant). The unitary $U$ slightly shifts $\left| m \right\rangle $ towards the center of the measurement spectrum, that is, $M_1 = \langle U S_x U^{\dagger} \rangle_m \approx  m(1-c^2 N^{-x}/2) $. At the same time, the second moment $M_2= \langle U S_x^2 U^{\dagger} \rangle_m$ increases because neighboring basis states are populated. This is quantified by the standard deviation $D = \sqrt{M_2 - M_1^2}$. A short calculation shows that $D \lesssim c N^{1-x/2}/(2 \sqrt{2})$ for any $m$. In particular, the case $x = 2$ gives $D \leq c/(2 \sqrt{2})$. In other words, the difference between $| m \rangle $ and $\left| \tilde{m} \right\rangle $ is limited to a small region within the spectrum. Therefore, a measurement resolution $\Delta \lesssim D = O(N^{1-x/2})$ is mandatory to resolve this difference. The larger the QFI and the shorter hence $t$, the larger are the requirements on the measurement resolution (see also Refs.~\cite{Gietka_Quantum-enhanced_2015,Oudot_Two-mode_2015}).

\textit{Modified scheme.---}
We are now in the position to present the main result of this paper. In order to counter the negative effect of unavoidable finite detector resolution, one can modify the measurement and increase its resolution by applying an additional unitary operation after $U$ (see Fig.~\ref{fig:scheme}). This is not possible with linear operators. However, similarly to the generation of $| \phi_0 \rangle $, one can apply another nonlinear operation $W$ between $U$ and the measurement. This means that we compare the two states $| \psi_0 \rangle = W V \left| 0 \right\rangle $ and $|\psi_1  \rangle = W U V \left| 0 \right\rangle $. Astonishingly, it turns out that $W$ can often be implemented using the same resources as the preparation step, as demonstrated now for two relevant examples.

\textit{Example 1: photonic squeezing.}  We start with the vacuum state $\left|  \mathrm{C} \right\rangle = \left| 0 \right\rangle $ and choose $V$ to be the standard squeezing operation $V = \exp[-\xi/2(a^{\dagger 2}-a^2)]$ with $\xi >0$. The state $| \phi_0 \rangle $ is hence squeezed along the $x$ axis. For the quadrature $X = 1/\sqrt{2}(a^{\dagger} + a)$, its distribution is a Gaussian with variance $V(X)=\frac{1}{2} e^{-2 \xi}$. We choose $H = P = -i/\sqrt{2}(a^{\dagger} - a)$, that is, the unitary displaces every state in phase space by $\alpha = t/\sqrt{2}$ along the $x$ axis. Then, the Bhattacharyya coefficient is $B_\Omega=\exp[-\frac{1}{2}\alpha^2/V(X)]$ by measuring $X$. For small enough $\alpha$ the bound \eqref{eq:3} witnesses a QFI which is inversely proportional to the variance and grows exponentially with the squeezing parameter. However coarse-graining decreases the distinguishability. With a similar model $\Omega_X(x)$ as in Eq.~(\ref{eq:5}) (replacing $S$ by $X$), the variance of $X$ increases to $V(X)= \frac{1}{2}e^{-2 \xi}+ \Delta^2$. This bounds the verifiable QFI to $\mathcal{B} \leq 1/\Delta^2$ regardless of the state.
Applying a second squeezing $W= \exp[\xi'/2(a^{\dagger 2}-a^2)]$  after the evolution $U$ allows to overcome this limitation. In the Heisenberg picture, one has $W^\dag X W = e^{\xi'} X$ and hence the spectrum is effectively stretched. In other words, the uncertainty from coarse-graining is effectively suppressed  $\Delta \rightarrow \Delta e^{-\xi'}$ [see also Fig.~\ref{fig:scheme} (b) for $\xi^{\prime} = \xi$].

\textit{Example 2: spin-squeezing.} For spins, analytic results are more difficult to obtain. However, as long as $\mu$ is not too large in Eq.~(\ref{eq:4}) (i.e., the state is not oversqueezed) approximate expressions for spin-squeezed states similar to the photonic squeezing can be found. Here, we numerically illustrate the impact of the back-squeezing operation $W = V^{\dagger}$ for $\mathcal{B}$ with the detector model (\ref{eq:5}) in Fig.~\ref{fig:optsqueezing}. We observe that the bound can be tight even for realistic $\Delta$ around ten with an improvement of up to two orders of magnitude compared to $W = \mathbbm{1}$. We observe very similar results in the oversqueezed regime.

\begin{figure}[htbp]
\centerline{\includegraphics[width=\columnwidth]{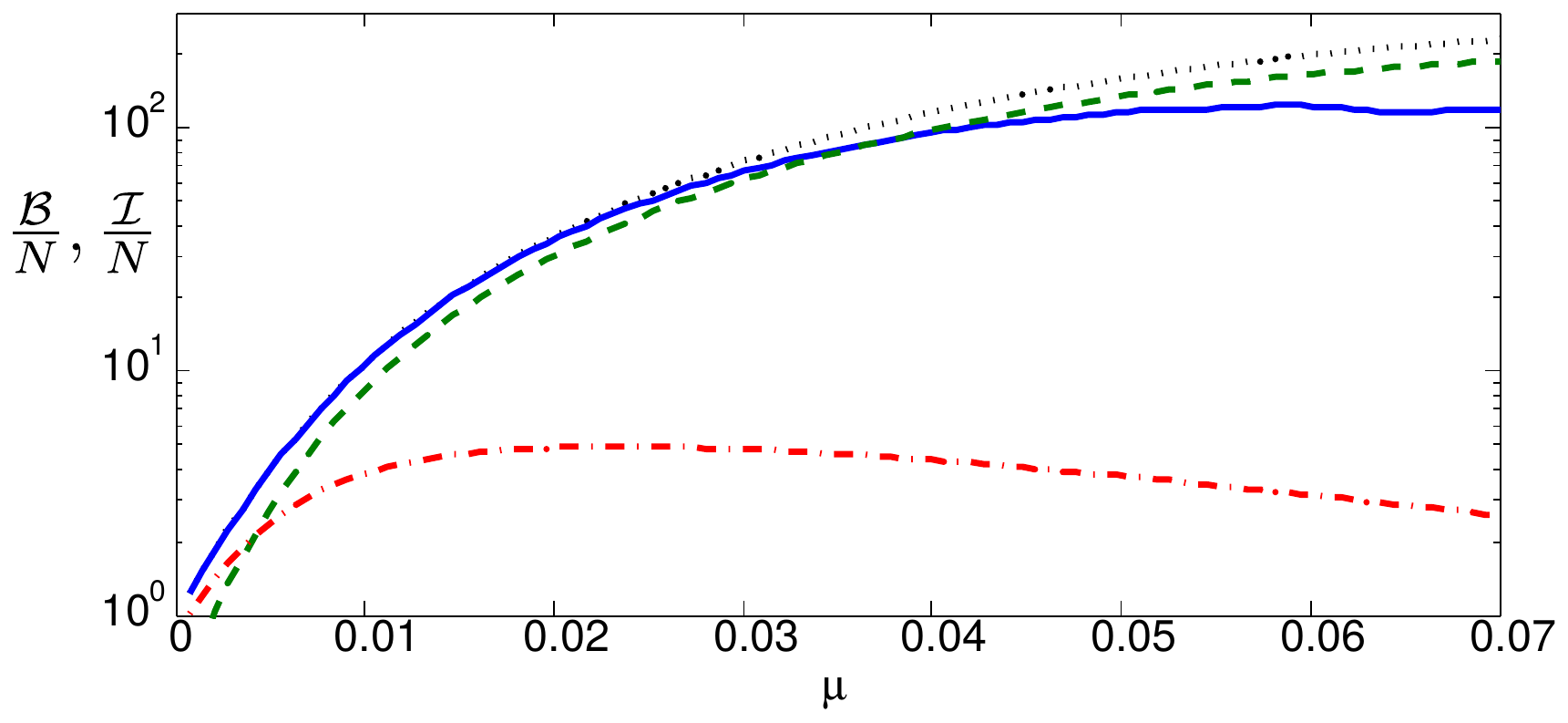}}
\caption[]{\label{fig:application} Lower bounds $\mathcal{B}/N \leq \mathcal{I}/N$ optimized over $\mu^{\prime}$ and $\nu^{\prime} $ in $W= \exp(-\frac{i}{2} \mu^{\prime} S_z^2) \exp(-i \nu^{\prime} S_x)$ for $N = 600$, $\Delta = 5$, $t = 10^{-3}$ and various $\mu$ parametrizing $| \mathrm{S}_{\mu} \rangle $ (blue solid line). The values are compared to $W = V^{\dagger}$ (green dashed line), $W = \mathbbm{1}$ (red dash-dotted line) and the true QFI $\mathcal{I}/N$ (black dotted line). We observe that for small $\mu$ the bound is tight and even surpasses the results for $W = V^{\dagger}$. In addition, a nontrivial $W$ is clearly advantageous in the oversqueezed regime, starting from $\mu \gtrsim 0.05$. }
\end{figure}

Although it seems that the choice $W = V^{\dagger}$ often leads to good results, it is not necessarily the case. On the other hand, we observe that one easily finds instances $W \neq V^{\dagger}$ that still help to overcome problems with detector resolution. This is particularly important in experiments where the sign of the interaction cannot be changed (see, e.g., Ref.~\cite{Riedel_Atom-chip-based_2010}). In Fig.~\ref{fig:application}, we present a numerical study for $\mathcal{B}$, where we search for operations $W$ composed of $ \exp(-\frac{i}{2} \mu^{\prime} S_z^2), \mu^{\prime} >0$, plus local rotations for different $| \mathrm{S}_{\mu} \rangle $ with $\mu^{\prime}$ generally different from $\mu$. For small squeezing, one can achieve even better results than for $W = V^{\dagger}$.

{ \textit{Connection to Leggett-Garg inequalities.---} The comparison of two states $| \psi_0 \rangle =W V \left| 0 \right\rangle $ and $|\psi_1  \rangle =W U V \left| 0 \right\rangle $ with coarse-grained collective measurements is reminiscent of the recently proposed NSIT conditions \cite{Kofler_Condition_2013}. These conditions are an alternative \cite{Clemente_Necessary_2015} to Leggett-Garg inequalities \cite{Leggett_Quantum_1985}, which aim to test so-called macro-realistic theories.
  The non-disturbance of (macroscopic) systems through measurements is the core assumption of such theories, which contrasts quantum mechanics. Therefore inequalities for correlations between sequential measurements derived from this assumption can be violated by a quantum system.

The simplest way to formulate the NSIT conditions is to witness the change in the measurement statistics of of some observable depending on whether or not another measurement has been preformed previously. Moreover, one is interested in measurements that have a meaningful ``macroscopic'' limit. For spins, collective measurements with a finite measurement resolution introduced in Eq.~\eqref{eq:5} are typical instances. 
Let us fix the first measurement to be along the $z$ axis, $\Omega_z(x)$, and the second to be along the $y$ axis, $\Omega_y(x)$. In between, we assume some time evolution $V$. (For the moment, $V$ is more general then in the first part of the paper.) The same unitary is applied to the system before the first measurement. In summary, one is interested in comparing the measurement statistics of the undisturbed state, $p(x) = \mathrm{Tr}[\rho_0 V^{\dagger} \Omega_y(x) V]$ where $\rho_0 = V \left| \mathrm{C}\right\rangle\!\left\langle \mathrm{C}\right| V^{\dagger}$, with an intermediately measured state averaged over all measurement results $z$, $q(x) = \mathrm{Tr}[\rho_{\mathrm{av}} V^{\dagger} \Omega_y(x) V]$, where $\rho_{\mathrm{av}} = \int dz \sqrt{\Omega_z(z)} \rho_0 \sqrt{\Omega_z(z)}$.
 Like before, the difference of $p(x)$ and $q(x)$ can be measured, for example, with the Bhattacharyya coefficient $B_{\Omega}$.

To see the connection to the QFI, notice that using the Fourier transform of $\sqrt{\Omega_z(z)}$ one can rewrite the averaged state as 
$\rho_{\mathrm{av}} = \int dk \sqrt{2/\pi} \Delta e^{-2 \Delta^2 k^2}  e^{-i k S_z}  \rho_0  e^{i k S_z}$.
 This compares to the previous situation comparing $| \psi_0 \rangle $ and $\left| \psi_1 \right\rangle $ if we take $W = V$ and $k = t$ and if $| \psi_1 \rangle $ is averaged over a Gaussian distribution.

In this context, it is interesting to note that it is not possible to violate Leggett-Garg inequalities with coarse-grained measurements (with $\Delta \gg \sqrt{N}$) if $V$ is generated by a linear Hamiltonian \cite{Kofler_Classical_2007}. The connection of the discussed NSIT condition to bound (\ref{eq:3}) for estimating the QFI from below gives two interesting insights. First, we observe that a large QFI $\mathcal{I}\equiv \mathcal{I}_{\rho_0}(S_z)$ is necessary to violate Leggett-Garg inequalities with coarse-grained detectors. Due to the concavity and positivity of the fidelity, Eq.~(\ref{eq:2}) can be rewritten for a mixed state $\rho_{\mathrm{av}}$
\begin{equation}
\label{eq:7}
\begin{split}
  F(\rho_0,\rho_{\mathrm{av}}) &\geq \int_{-\pi/
    \sqrt{\mathcal{I}}}^{\pi/ \sqrt{\mathcal{I}}} dk\frac{\sqrt{2}\Delta}{\sqrt{\pi}} e^{-2 \Delta^2 k^2}
  \cos \left(\frac{1}{2}\sqrt{\mathcal{I}}\, k \right)\\ 
& \geq e^{-\frac{\mathcal{I}}{32 \Delta^2}} -  \text{Erfc}\Big( \frac{\sqrt{2} \pi  \Delta }{\sqrt{\mathcal{I}}}\Big).
\end{split}
\end{equation}
Therefore, given a large measurement uncertainty, say $\Delta = \sqrt{N}$, quantum states with a large QFI $\mathcal{I} \gg N$ are necessary to violate NSIT conditions and hence Leggett-Garg inequalities. On the other hand, with the results presented previously, we find that $V$ generated by a quadratic Hamiltonian is sufficient to overcome the limitations of a coarse-grained detection device. Given that these operators are nowadays routinely implemented in the lab, strong violations of  Leggett-Garg inequalities in mesoscopic system sizes seems to be within reach.

\textit{Summary and outlook.---}
In this paper, we proposed a simple protocol for the experimental verification of large QFI even with coarse-grained measurements. With an additional squeezing operation right before the measurement, present-day spin-squeezing experiments could increase lower bounds on the QFI by up to two orders of magnitude. Notably, it is not necessary to impose $W = V^{\dagger}$ to achieve very good results. This paves the way for a reliable detection of large scale QFI and multipartite entanglement using collective measurements only.

At the same time, it provides an accessible scheme for quantum metrology, where the action of the unitary $U$ is determined by a (partially) unknown parameter. It is expected that the proposed application of $W$ also helps to improve the sensitivity in these scenarios. 
It is an open but appealing question whether a well chosen $W$ helps to reveal other nonclassical quantities in multipartite systems.

\textit{Acknowledgments.---} We thank \v Caslav Brukner, Roman Schmied, Nicolas Sangouard and Nicolas Gisin for stimulating discussions. This work was supported by the National Swiss Science Foundation (SNSF), Grant number PP00P2-150579 and  P2GEP2-151964, the Austrian Science Fund (FWF), Grant numbers J3462, P24273-N16, P28000-N27, SFB F40-FoQus F4012-N16, the COST Action No.
MP1006 and the European Research Council (ERC MEC).

\textit{Note added:} While preparing this paper, we became aware of Ref.~\cite{Davis_Approaching_2015}, which demonstrates how to restore Heisenberg limit in quantum metrology protocols with the help of $W = V^{\dagger}$ in the presence of coarse-grained collective observables.

\bibliographystyle{apsrev4-1}
\bibliography{DetectingLargeQFI}

\end{document}